\newlength{\absize}
\documentclass[12pt]{article}

\usepackage{epstopdf}
\usepackage{epsfig}
\usepackage{amsmath,amsthm, amsfonts,amssymb} 

\usepackage{rotating}
\usepackage{verbatim}
\usepackage{longtable}
\usepackage{lscape}
\usepackage{subfig}
\usepackage{float}
\usepackage{color,soul,mathtools} 
\usepackage[dvipsnames]{xcolor} 
\usepackage[all]{xy} 
\usepackage{braket} 
\usepackage{graphicx} 
\usepackage{indentfirst} 
\usepackage{bm} 
\usepackage{mathrsfs} 
\usepackage{latexsym} 
\usepackage{hyperref} 
\hypersetup{
	colorlinks = true,
	linkcolor = blue,
	urlcolor = blue,
	citecolor = blue,
	anchorcolor = black,
}
\usepackage{microtype} 
\usepackage{bookmark} 
\usepackage{tikz}[2010/10/13] 

\usepackage{mathtools} 
\usepackage{ifpdf} 
\usepackage{doi} 

\usepackage{url}
\usepackage{rotating}
\usepackage[innercaption]{sidecap}
\usepackage{pdfpages}
\usepackage{xspace}

\usepackage{fancyhdr}
\newtheorem{Thm}{Theorem}[section]

\newtheorem{lemma}[Thm]{\bf Lemma}

\newtheorem{theorem}[Thm]{\bf Theorem}

\newcommand{\be}{\begin{equation}}
\newcommand{\ee}{\end{equation}}

\newcommand{\beq}{\begin{eqnarray}}
\newcommand{\eeq}{\end{eqnarray}}
\newcommand{\beqs}{\begin{eqnarray*}}
	\newcommand{\eeqs}{\end{eqnarray*}}

\begin{document}

\thispagestyle{empty}
\pagestyle{empty}
\newcommand{\starttext}{\newpage\normalsize
	\pagestyle{plain}
	\setlength{\baselineskip}{3ex}\par
	\setcounter{footnote}{0}
	\renewcommand{\thefootnote}{\arabic{footnote}}
}
\newcommand{\preprint}[1]{\begin{flushright}
		\setlength{\baselineskip}{3ex}#1\end{flushright}}
\renewcommand{\title}[1]{\begin{center}\LARGE
		#1\end{center}\par}
\renewcommand{\author}[1]{\vspace{2ex}{\large\begin{center}
			\setlength{\baselineskip}{3ex}#1\par\end{center}}}
\renewcommand{\thanks}[1]{\footnote{#1}}
\renewcommand{\abstract}[1]{\vspace{2ex}\normalsize\begin{center}
		\centerline{\bf Abstract}\par\vspace{2ex}\parbox{\absize}{#1
			\setlength{\baselineskip}{2.5ex}\par}
\end{center}}

\title{An Algebraic Framework for Multi-Qudit Computations with Generalized Clifford Algebras}
\author{
	Robert~Lin
	\\ \medskip
	Department of Physics \\
	Harvard University \\
	Cambridge, MA 02138
}
\date{\today}

\abstract{In this article, we develop an algebraic framework of axioms which abstracts various high-level properties of multi-qudit representations of generalized Clifford algebras. We further construct an explicit model and prove that it satisfies these axioms. Strengths of our algebraic framework include the minimality of its assumptions, and the readiness by which one may give an explicit construction satisfying these assumptions. In terms of applications, this algebraic framework provides a solid foundation which opens the way for developing a graphical calculus for multi-qudit representations of generalized Clifford algebras using purely algebraic methods, which is addressed in a follow-up paper.}

\newpage
\starttext

\section{Introduction}

The following physics question motivates this article: Can we learn new things about quantum entanglement by studying a purely algebraically formulated graphical calculus for the generalized Clifford algebras\footnote{The earliest paper introducing generalized Clifford algebras appears to be \cite{Morinaga} in 1952. Other early work included \cite{Yamazaki} in 1964, \cite{Pop} in 1966, and \cite{Morris} in 1967.}? 	Whereas much effort has been expended on studying multiqubit settings via the Clifford algebras, also called Majorana fermions \cite{TopoMajorana}, the corresponding setting for multiple qudits, the generalized Clifford algebras, has been much less studied \cite{Cobanera}. In this setting, the work of \cite{Cobanera} and \cite{JL} studied braiding operators defined using the generalized Clifford algebra which are unitary operations that entangle neighboring qudits (multi-dimensional vector spaces). Thus, when one applies a sequence of braiding operators to the ground (or vacuum) state, we expect different kind of entangled states to result, depending on the sequence and on the braidings in the sequence. Some questions which are natural to consider are the following: Is there an easy way to classify the resulting kinds of entanglement using the graphical calculus? How does the classification depend on the number of qudits involved?

 In terms of these questions, a  graphical calculus for the generalized Clfford algebras which can be logically motivated \textit{and} carried out relatively simply using purely algebraic manipulations is currently lacking in the following sense. To be precise, here we consider the bar for ``simplicity" of algebraic representation and manipulation as one which could be learned by a computer algebra system which can be trained to carry out typical calculations and simplifications which arise in proofs, or what is nowadays called \textit{artificial intelligence}. Thus, one motivation for this work is to work out a simple algebraic version of the theory of \cite{JL}, which does not depend on higher algebraic notions such as planar algebras, and hence would form a logically compact and consistent framework for computer algebra manipulation\footnote{An important Mathematica package which one might use to resolve this problem from the calculational perspective at the level of the generalized Clifford algebra, but not manifestly at the level of the representation, is NCAlgebra \cite{Helton}, developed by Bill Helton and collaborators. We briefly comment here that by using projection operators combined with the axioms we introduce, one ought to be able to extend the domain of application to the level of vectors as well.
 }. Such a \textit{simply algebraically formulated} graphical calculus would be useful as machinery for proofs and new derivations of a purely algebraic type.

To get answers to these questions in a systematic manner, we have developed an algebraic framework which abstracts two main features, the existence of a distinguished ground state, and the basis property of the generalized Clifford algebras acting on this ground state. Our abstraction clarifies the algebraic dependencies involved in working with the vector representation, as opposed to working purely in the generalized Clifford algebras, and can be considered to provide a simple conceptual algebraic basis for the more elaborate machinery employed in \cite{JL}. Because our motivation is somewhat specific, being to strive for a minimal set of axioms (and minimal additional machinery) from which fairly general results follow, our approach differs from other related works on generalized Clifford algebras \cite{Cobanera}\cite{JP}. In terms of applications, this algebraic framework provides a solid foundation which opens the way for developing a graphical calculus for multi-qudit representations of generalized Clifford algebras using purely algebraic methods. Furthermore, our algebraic framework allows us to resolve an open question in \cite{Cobanera}, regarding proving algebraically a general solution to the Yang-Baxter equation (which establishes the braiding nature). Both will be addressed in a follow-up article. 

As a general comment, while the algebraic manipulations involved are essentially the well-known Jordan-Wigner transformation, our work brings to the fore in a \textit{unified} manner certain aspects (such as the necessity of justifying the choice of square root of the $N$th-root of unity \textit{at the same time} that one verifies the properties of the unitary representation, for proper rigor) and clarifies the necessity of a distinguished ground state to establish algebraic properties of general usefulness.

\section{The Algebraic Framework}

Fix $N$ a positive integer greater than 1, $n$ a positive integer at least 1, and consider the \textbf{generalized Clifford algebra} $\mathcal{C}_{2n}^{(N)}$ generated by $c_1$, $c_2$, $c_3$, $\ldots$ , $c_{2n}$ subject to $c_i c_j = q c_j c_i$ if $i<j$, and $c_i^N=1$ for all $i$. Here, $q=\exp(2\pi i/N)$ is a primitive Nth root of unity. When $N=2$, one recovers the Clifford algebra with $2n$ generators.

Whereas previous authors \cite{JL} have considered fairly elaborate frameworks for working with the generalized Clifford algebras in diagrammatic fashion, I have found that the following axiomatization gives rise to a very straightforward algebraic framework for a graphical calculus for the generalized Clifford algebras.

\textbf{Axiom 1}:
Let $\mathcal{V}^{N^n}(\mathbb{C})$ be a complex vector space upon which the generalized Clifford algebra is realized as unitary $N^n$ by $N^n$ matrix operators. Assume that there exists a state (which we call the ground state) which is a tensor of states $\ket{\Omega}$, $\ket{\Omega}^{\otimes n}$, that satisfies the following algebraic identity:
$$ c_{2k-1} \ket{\Omega}^{\otimes n} = \zeta\, c_{2k} \ket{\Omega}^{\otimes n}$$
for all $k=1,2,\ldots, n$, where $\zeta$ is a square root of $q$ such that $\zeta^{N^2}=1$. 

In addition, for each qudit, the projector $E_{k}$ onto the $k$th qudit's ground state $\ket{\Omega}$ is assumed to satisfy 
$$ c_{2k-1} E_{k} = \zeta\, c_{2k} E_{k}.$$

\textbf{Axiom 2: Scalar product}:
The set $\{c_2^{a_1} c_4^{a_2} \ldots c_{2n}^{a_n} \ket{\Omega}^{\otimes n}: a_i=0,1, \ldots, N-1\}$ is an orthonormal basis for $\mathcal{V}^{N^n}(\mathbb{C})$.

By associativity, all algebraic manipulations can be reduced to pairwise operations, and it does not matter the order in which we perform simplifications.

We will show that these two axioms can be simultaneously satisfied by giving an explicit construction and verifying that this construction verifies all the properties and assumptions given in the axioms. In other words, the axiomatic framework is not a vacuous one. This is important since it implies that all results derived \textit{from} the axioms are true in at least one explicit model.

The easiest way to construct a unitary representation of the generalized Clifford algebra satisfying the above axioms is to work backward from the assumption that the axioms hold, and to calculate the action of the generalized Clifford algebra on the basis states given by $\{c_2^{a_1} c_4^{a_2} \ldots c_{2n}^{a_n} \ket{\Omega}^{\otimes n}: a_i=0,1, \ldots, N-1\}$. For convenience, we label these basis states by the tuples $\ket{a_1,a_2,\ldots,a_n}$. Then $c_{2k}\ket{a_1,a_2,\ldots,a_n} = c_{2k}c_2^{a_1} c_4^{a_2} \ldots c_{2n}^{a_n} \ket{\Omega}^{\otimes n}  = q^{-\sum_{i<k} a_i} c_{2}^{a_1} c_4^{a_2} \cdots c_{2k}^{a_k+1} \cdots c_{2n}^{a_{n}}\ket{\Omega}^{\otimes n}=q^{-\sum_{i<k} a_i} \ket{a_1,a_2,\ldots, a_k+1,\ldots,a_n}$. Thus, we now \textbf{define} $c_{2k}$ as a matrix operator on the basis $\ket{a_1,a_2,\ldots,a_n}$ via 
$$c_{2k}\ket{a_1,a_2,\ldots,a_n}:= q^{-\sum_{i<k} a_i} \ket{a_1,a_2,\ldots, a_k+1,\ldots,a_n}$$
for $a_i=0,1,\ldots, N-1$.

Now, let's calculate the action of $c_{2k-1}$ on this same basis. We first need to find $\zeta$ such that $\zeta^2=q$ and $\zeta^{N^2}=1$. 

\begin{lemma}
	\label{zeta}
	Let $q=\exp(2\pi i/N)$. If $N$ is odd, $\zeta=-\exp(\pi i/N)$ is the only square root of $q$ satisfying $\zeta^{N^2}=1$. If $N$ is even, setting $\zeta$ to be either square root of $q$ will satisfy $\zeta^{N^2}=1$.
\end{lemma}
\begin{proof}
	$q=e^{i\frac{2\pi}{2N+1}}$, $\zeta =\pm e^{i\frac{\pi}{2N+1}}$ for odd case yields $\zeta^{(2N+1)^2} = \pm \exp(i\pi (2N+1))=(\pm 1)(-1)=\mp 1$, so one chooses the $-$ sign. For even case,  $q=e^{i\frac{2\pi}{2N}}=e^{i\pi/N}$, then $\zeta =\pm e^{i\pi/2N}\rightarrow \zeta^{(2N)^2}=\zeta^{4N^2}=(\pm e^{i\pi/2N})^{4N^2}=e^{i\pi (2N)}=1$. 
\end{proof}

Thus, we choose $\zeta$ according to the lemma \ref{zeta}. Now using the axioms and applying $c_{2k-1}$ to the basis elements yields
$c_{2k-1} \ket{a_1,a_2,\ldots,a_n} =c_{2k-1}\, c_2^{a_1} c_4^{a_2} \ldots c_{2n}^{a_n} \ket{\Omega}^{\otimes n} = \zeta q^{a_k-\sum_{i<k} a_i} c_2^{a_1} c_4^{a_2} \cdots c_{2k}^{a_k+1}\ldots c_{2n}^{a_n} \ket{\Omega}^{\otimes n}$, which then gives $\zeta \,q^{a_k-\sum_{i<k} a_i}\ket{a_1,a_2,\ldots,a_{k}+1\,\ldots, a_{n}} $. So we \textbf{define}
$$c_{2k-1}\ket{a_1,a_2,\ldots,a_n}:= \zeta \, q^{a_k} q^{-\sum_{i<k} a_i} \ket{a_1,a_2,\ldots, a_k+1,\ldots,a_n}$$

We are now in a position to state the following theorem.

\begin{theorem}
	Consider an orthonormal basis of the complex vector space $\mathcal{V}^{N^n}(\mathbb{C})$ labeled by the tuples $(a_1,a_2,\ldots, a_n)$, for $a_i=0,1,\ldots, N-1$, i.e. the states are given by $\ket{a_1,a_2,\ldots, a_{n}}$. We can identify this complex vector space with a tensor of $n$ $N$-dimensional complex vector spaces such that $\ket{a_1,a_2,\ldots,a_n}=\ket{a_1}\otimes \cdots \otimes \ket{a_n}$.
	
	Define the matrix operators $c_{2k-1}$, $c_{2k}$ by their action on the orthonormal basis $\ket{a_1,a_2,\ldots, a_{n}}$ via
	$$c_{2k}\ket{a_1,a_2,\ldots,a_n}:= q^{-\sum_{i<k} a_i} \ket{a_1,a_2,\ldots, (a_k+1)(\text{mod }N),\ldots,a_n}$$
	and
	$$c_{2k-1}\ket{a_1,a_2,\ldots,a_n}:= \zeta \, q^{a_k} q^{-\sum_{i<k} a_i} \ket{a_1,a_2,\ldots, (a_k+1) (\text{mod }N),\ldots,a_n}$$
	for all $k=1,2,\ldots, n$, where $\zeta$ is chosen according to the lemma \ref{zeta}.\footnote{For convenience, we will omit all the mod N qualifiers, and simply identify states with the same indices mod N. This identification is justified since the coefficients of $q$ to some power are invariant under shifts of the indices mod N.}
	
	Define the matrix operators $E_k$, for $k=1,2,\ldots,n$ by the linear extension of their action on the orthonormal basis via
	$$
	E_k \ket{a_1,a_2,\ldots, a_k,\ldots,a_n} = \delta_{a_k,0}\ket{a_1,a_2,\ldots, 0,\ldots, a_n}	
	$$	
	for all $a_i=0,1,\cdots, N-1$, $i=1,2,\cdots,n$.
	
	Define the ground state $$\ket{\Omega}:=\ket{0}$$
	so that
	$$\ket{\Omega}^{\otimes n}:=\ket{0,0,\cdots,0}$$
	
	Then the matrix operators $c_{2k-1}$, $c_{2k}$, $E_k$ and the ground state $\ket{\Omega}$ satisfy axioms 1 and 2.	
\end{theorem}
\begin{proof}
	First, we need to show that $c_{2k-1}$, $c_{2k}$ are unitary, and that $c_{2k-1}^{N}=1$, $c_{2k}^{N}=1$, as well as $c_{i}c_{j}=q c_{j} c_{i}$ for $i<j$.
	
	Unitarity can be shown by showing that $c_{2k}^{\dagger}\, c_{2k}=c_{2k}c_{2k}^{\dagger}=1$. Note that the dagger operation is just the usual conjugate transpose operation in the orthonormal basis setting.
	
	$$c_{2k}= \sum_{a_j=0,1,\ldots,N-1} q^{-\sum_{i<k} a_i} \ket{a_1,a_2,\ldots, a_k+1,\ldots,a_n}\bra{a_1,a_2,\ldots,a_n}$$
	implies 
	$$c_{2k}^{\dagger}= \sum_{a_j=0,1,\ldots,N-1} q^{\sum_{i<k} a_i} \ket{a_1,a_2,\ldots,a_n}\bra{a_1,a_2,\ldots, a_k+1,\ldots,a_n}$$
	so clearly the outcome is
	$$
	c_{2k} c_{2k}^{\dagger} = \sum_{a_j=0,1,\ldots,N-1} \ket{a_1,a_2,\ldots, a_k+1,\ldots,a_n}\bra{a_1,a_2,\ldots, a_k+1,\ldots,a_n} = 1
	$$
	and that
	$$
	c_{2k}^{\dagger}c_{2k}  = \sum_{a_j=0,1,\ldots,N-1} \ket{a_1,a_2,\ldots, a_k,\ldots,a_n}\bra{a_1,a_2,\ldots, a_k,\ldots,a_n} = 1.
	$$
	
	Similarly, for $c_{2k-1}$, we have that
	$$
	c_{2k-1}= \sum_{a_j=0,1,\ldots,N-1} \zeta \, q^{a_k} q^{-\sum_{i<k} a_i} \ket{a_1,a_2,\ldots, a_k+1,\ldots,a_n}\bra{a_1,a_2,\ldots,a_n}
	$$
	and
	$$
	c_{2k-1}^{\dagger}= \sum_{a_j=0,1,\ldots,N-1} \zeta^{-1} \, q^{-a_k} q^{\sum_{i<k} a_i} \ket{a_1,a_2,\ldots,a_n} \bra{a_1,a_2,\ldots, a_k+1,\ldots,a_n}
	$$
	implying that
	$$
	c_{2k-1}\, c_{2k-1}^{\dagger}= \sum_{a_j=0,1,\ldots,N-1} \ket{a_1,a_2,\ldots,a_k+1, \ldots, a_n} \bra{a_1,a_2,\ldots, a_k+1,\ldots,a_n}=1.
	$$
	And also that 
	$$
	c_{2k-1}^{\dagger} c_{2k-1} = \sum_{a_j=0,1,\ldots,N-1} \ket{a_1,a_2,\ldots,a_k, \ldots, a_n} \bra{a_1,a_2,\ldots, a_k,\ldots,a_n}=1
	$$
	This concludes the check for \textbf{unitarity}.
	
	For the relations satisfied by $c_{2k}$, and $c_{2k-1}$, we have that
	since $$c_{2k}\ket{a_1,a_2,\ldots,a_n}= q^{-\sum_{i<k} a_i} \ket{a_1,a_2,\ldots, a_k+1,\ldots,a_n}$$ implies that $$c_{2k}^{N}\ket{a_1,a_2,\ldots,a_n}= q^{-N\sum_{i<k} a_i} \ket{a_1,a_2,\ldots, a_k+N,\ldots,a_n} =\ket{a_1,a_2,\ldots, a_k,\ldots,a_n}, $$ it follows by unique linear extension that $c_{2k}^{N}=1$.
	
	The statement for $c_{2k-1}$ is a bit more involved to show. Starting from $c_{2k-1}\ket{a_1,a_2,\ldots,a_n}= \zeta \, q^{a_k} q^{-\sum_{i<k} a_i} \ket{a_1,a_2,\ldots, a_k+1,\ldots,a_n}$, we obtain that
	$$c_{2k-1}^2\ket{a_1,a_2,\ldots,a_n}= \zeta^2 \, q^{a_k+(a_k+1)} q^{-2\sum_{i<k} a_i} \ket{a_1,a_2,\ldots, a_k+2,\ldots,a_n},$$ so that $c_{2k-1}^3\ket{a_1,a_2,\ldots,a_n}= \zeta^3 \, q^{a_k+(a_k+1)+(a_k+2)} q^{-3\sum_{i<k} a_i} \ket{a_1,a_2,\ldots, a_k+3,\ldots,a_n}$, and inductively, one obtains that $$c_{2k-1}^m\ket{a_1,a_2,\ldots,a_n}= \zeta^m \, q^{m\, a_k+ (0+1+\cdots +(m-1))} q^{-m\sum_{i<k} a_i} \ket{a_1,a_2,\ldots, a_k+m,\ldots,a_n}.$$
	
	Plugging in $m=N$, we get that 
	\begin{align}
	c_{2k-1}^N\ket{a_1,a_2,\ldots,a_n}&:= \zeta^N \, q^{N\, a_k+ N(N-1)/2} q^{-N\sum_{i<k} a_i} \ket{a_1,a_2,\ldots, a_k,\ldots,a_n} \\
	&=\zeta^N \, q^{(N^2-N)/2}\ket{a_1,a_2,\ldots, a_k,\ldots,a_n}.
	\end{align}
	Now things get interesting. If $N$ is even, $\zeta = \pm q^{1/2}$ implies that $\zeta^N =q^{N/2}$, in which case $q^{N/2}q^{(N^2-N)/2}=q^{N^2/2}=1$. If $N$ is odd, $\zeta = - q^{1/2}$ implies that $\zeta^N = - q^{N/2}$, in which case $- q^{N/2} q^{(N^2-N)/2} = - q^{N^2/2} = -e^{(2 \pi i/N) \cdot(N^2/2)}=-e^{N \pi i} = - (-1)^N = - (-1)=1$! So we have shown that $c_{2k-1}^N=1$.
	
	To show $c_i c_j = q c_j c_i$ for all $i<j$, observe that
	$$c_{2k}\ket{a_1,a_2,\ldots,a_n}= q^{-\sum_{i<k} a_i} \ket{a_1,a_2,\ldots, a_k+1,\ldots,a_n}$$
	yields
	$$c_{2l} c_{2k}\ket{a_1,a_2,\ldots,a_n}= q^{-\sum_{i<l}a_i} q^{-\sum_{i<k} a_i} \ket{a_1,a_2,\ldots, a_l+1,\ldots, a_k+1,\ldots,a_n}$$
	if $l<k$. Meanwhile, $c_{2l}\ket{a_1,a_2,\ldots,a_n}= q^{-\sum_{i<l} a_i} \ket{a_1,a_2,\ldots, a_l+1,\ldots,a_n}$ yields
	$$c_{2k} c_{2l}\ket{a_1,a_2,\ldots,a_n}= q^{-(\sum_{i<k} a_i)-1}q^{-\sum_{i<l} a_i} \ket{a_1,a_2,\ldots, a_l+1,\ldots,a_k+1,\ldots,a_n}.$$
	So $c_{2k} c_{2l} = q^{-1} c_{2l} c_{2k}$, i.e. $$ c_{2l} c_{2k} = q\, c_{2k} c_{2l}$$
	for $l<k$.
	
	Repeating the procedure for $c_{2k-1}$, $c_{2l-1}$, we get that for $l<k$:
	$c_{2k-1}\ket{a_1,a_2,\ldots,a_n}= \zeta \, q^{a_k} q^{-\sum_{i<k} a_i} \ket{a_1,a_2,\ldots, a_k+1,\ldots,a_n}$ implies 
	$$c_{2l-1} c_{2k-1}\ket{a_1,a_2,\ldots,a_n}= \zeta^2 \, q^{a_k} q^{a_l} q^{-\sum_{i<k} a_i} q^{-\sum_{i<l} a_i}\ket{a_1,a_2,\ldots, a_l+1,\ldots, a_k+1,\ldots,a_n}$$ but swapping the order leads to 
	$$c_{2k-1} c_{2l-1}\ket{a_1,a_2,\ldots,a_n}= \zeta^2 \, q^{a_k} q^{a_l} q^{-\sum_{i<k} a_i} q^{-\sum_{i<l} a_i}q^{-1}\ket{a_1,a_2,\ldots, a_l+1,\ldots, a_k+1,\ldots,a_n}$$ since the $c_{2k-1}$ notices that the index on the $l$ qudit has been increased by 1. Thus, $c_{2k-1} c_{2l-1} = q^{-1} c_{2l-1} c_{2k-1}$, i.e.
	$$c_{2l-1} c_{2k-1} = q c_{2k-1} c_{2l-1}$$
	for $l<k$.
	
	Meanwhile, for $c_{2k-1}$ and $c_{2k}$, we have that
	$$c_{2k-1} c_{2k} \ket{a_1,a_2,\ldots, a_n} = \zeta\, q^{a_k +1} q^{-2\sum_{i<k} a_i} \ket{a_1,a_2,\ldots, a_{k}+2,\ldots, a_{n}}$$
	$$ c_{2k} c_{2k-1} \ket{a_1,a_2,\ldots,a_{n}} = \zeta q^{a_k} q^{-2\sum_{i<k} a_i} \ket{a_1,a_2,\ldots,a_{k} +2,\ldots, a_{n}}$$
	so $$c_{2k-1} c_{2k} = q c_{2k} c_{2k-1}.$$
	
	For $c_{2l-1}$, $c_{2k}$, with $l<k$, we have that
	$$c_{2l-1} c_{2k} \ket{a_1,a_2,\ldots, a_n} = \zeta q^{a_l} q^{-\sum_{i<l} a_i} q^{-\sum_{i<k} a_i} \ket{a_1,a_2,\ldots,a_{l}+1,\ldots, a_{k} +1,\ldots, a_{n}}$$
	whereas
	$$c_{2k} c_{2l-1} \ket{a_1,a_2,\ldots, a_n} = \zeta q^{a_l} q^{-\sum_{i<l} a_i} q^{-\sum_{i<k} a_i} q^{-1}\ket{a_1,a_2,\ldots,a_{l}+1,\ldots, a_{k} +1,\ldots, a_{n}}$$
	since $c_{2k}$ notices the change in the index of the lth qudit. So
	$$
	c_{2l-1} c_{2k} = q\, c_{2k} c_{2l-1}
	$$	for $l<k$.
	
	Finally, for $c_{2l}$, $c_{2k-1}$ with $l<k$, we have that
	$$c_{2l} c_{2k-1} \ket{a_1,a_2,\ldots,a_n}=\zeta q^{a_k} q^{-\sum_{i<k} a_i} q^{-\sum_{i<l} a_i} \ket{a_1,a_2,\ldots,a_l+1,\cdots, a_k+1,\cdots,a_n} $$
	$$c_{2k-1} c_{2l} \ket{a_1,a_2,\ldots,a_n} = \zeta q^{a_k} q^{-\sum_{i<k} a_i} q^{-\sum_{i<l} a_i} q^{-1} \ket{a_1,a_2,\ldots,a_l+1,\cdots, a_k+1,\cdots,a_n}$$
	so 
	$$c_{2l} c_{2k-1} = q c_{2k-1} c_{2l}
	$$for $l<k$.
	
	The above calculations showed that we have constructed a unitary representation of the generalized Clifford algebra. Now we have to show the other aspects of axiom 1 are true as well.
	
	For the algebraic identity for $c_{2k-1}$, $c_{2k}$, and the ground state, we have that
	$$c_{2k-1}\ket{0,0,\ldots,0}=\zeta \ket{0,0,\ldots,0,1,0,\ldots,0}$$
	and
	$$c_{2k}\ket{0,0,\ldots,0}=\ket{0,0,\ldots,0,1,0,\ldots,0}$$
	with the 1 appearing on the kth qudit. Thus,
	$$c_{2k-1} \ket{0,0,\ldots,0} = \zeta c_{2k} \ket{0,0,\ldots,0}.$$
	
	For the algebraic identity involving $c_{2k-1}$, $c_{2k}$, $E_k$, we have
	$$c_{2k-1} E_k \ket{a_1,a_2,\ldots,a_{k-1},a_k,a_{k+1},\ldots,a_n} = \zeta q^{-\sum_{i<k} a_i} \delta_{a_k,0}  \ket{a_1,a_2,\ldots,a_{k-1},1,a_{k+1},\ldots,a_n}$$
	$$c_{2k} E_k \ket{a_1,a_2,\ldots,a_{k-1},a_k,a_{k+1},\ldots,a_n} = q^{-\sum_{i<k} a_i} \delta_{a_k,0}  \ket{a_1,a_2,\ldots,a_{k-1},1,a_{k+1},\ldots,a_n}.$$ Thus, 
	$$c_{2k-1} E_k = \zeta c_{2k} E_k$$
	for all $k=1,2,3,\ldots,n$. 
	
	This concludes the proof that axiom 1 is satisfied.
	
	To show axiom 2 is satisfied, i.e. that the set $\{c_2^{a_1} c_4^{a_2} \ldots c_{2n}^{a_n} \ket{0}^{\otimes n}: a_i=0,1, \ldots, N-1\}$ is an orthonormal basis for $\mathcal{V}^{N^n}(\mathbb{C})$, it suffices to note that each power of $c_{2k}$ raises the kth index by 1 and multiplies the state by a complex number of modulus 1. Thus, up to phase factors, the set $\{c_2^{a_1} c_4^{a_2} \ldots c_{2n}^{a_n} \ket{0}^{\otimes n}: a_i=0,1, \ldots, N-1\}$ is the same as $\{\ket{a_1,a_2,\ldots, a_n}:a_i=0,1,\ldots,N-1\}$, which by construction is an orthonormal basis for $\mathcal{V}^{N^n}(\mathbb{C})$. Hence, the set $\{c_2^{a_1} c_4^{a_2} \ldots c_{2n}^{a_n} \ket{0}^{\otimes n}: a_i=0,1, \ldots, N-1\}$ is an orthonormal basis as well.

\end{proof}

\subsection{Discussion}

This paper gives an axiomatic framework for an entirely algebraic approach to doing computation with multiple qudits using the generalized Clifford algebras. Since the application of generalized Clifford algebras to quantum computation is still in a nascent state, relative to the development of quantum computation schemes using Clifford algebras, i.e. Majorana fermions \cite{TopoMajorana}, it is envisioned that such an axiomatic framework will be of general utility in developing quantum computation of similar interest in the multiqudit setting.

Our major technical achievement is the abstraction of various concrete operational properties into high-level statements which 
highlight  
the nontrivial algebraic relations satisfied by the ground state and the projection operators under ``local" actions of the generalized Clifford algebra. Furthermore, the axiomatization emphasizes the particularly rigid structure imposed by the scalar product. Intuitively, this abstraction is a very appealing result and realizes in a purely algebraic manner certain themes of \cite{JL}, which appears to be related to standard themes in quantum error correction, in particular the stabilizer formalism of Gottesman \cite{Gottesman-Clifford}. One notable difference is that the operators involved in Gottesman's stabilizer formalism commute; in our case they do not, and are thus more in the spirit of the Majorana operators of \cite{TopoMajorana}, which are a special case of generalized Clifford algebras when $N=2$.

In terms of physics, one may think of these algebraic identities as corresponding to the introduction of internal structure, in the sense of the particle physics mantra that if particles are not point particles (e.g., possessing spin or other quantum numbers), their internal dynamics can be illuminated by scattering experiments\footnote{For example, deep inelastic scattering experiments showed that baryons had substructure  \cite{Friedman}. On the theoretical side, Bjorken was using current algebra \cite{Bjorken} to obtain  scaling laws (for the differential cross section in terms of the momentum transfer)  which concorded with these experimental results.}. It seems reasonable to take such analogies more seriously in light of the recent scattering experiments with anyons \cite{Bartolomei}.

At a mathematical level, the familiar adage is that ``more structure equals more ease of computation"; at the same time, the more structure one has, the harder it becomes to verify that the resulting theory is a consistent one. Although one can work abstractly, the abstract proofs of consistency may not be accessible to physicists, at least not at a level at which he or she would be comfortable verifying. Thus, it is desirable to present explicit constructions of models satisfying an axiomatic theory, such as the one presented in this paper.

\section{Acknowledgments}

I acknowledge my intellectual debt to Professor Howard Georgi, for helpful discussions on the importance of working with concrete vector representations, during a reading and research course I took with him in spring 2018. I also express my gratitude to Professor Peter Shor for a very incisive question he asked, which was very helpful to me in formulating the final axioms.

I wish to thank my advisor, Professor Arthur Jaffe, for the vibrant research environment he has provided. I have been supported during the writing of this paper by ARO grant W911NF-20-1-0082 through the MURI project ``Toward Mathematical Intelligence and Certifiable Automated Reasoning: From Theoretical Foundations to Experimental Realization."

\end{document}